\documentclass[prd, twocolumn, nofootinbib]{revtex4}

\usepackage{epsfig} 

\newcommand{\beq}{\begin{equation}}
\newcommand{\eeq}{\end{equation}}
\newcommand{\beqa}{\begin{eqnarray}}
\newcommand{\eeqa}{\end{eqnarray}}
\newcommand{\om}{\Omega_m}
\newcommand{\ow}{\Omega_w}
\newcommand{\ome}{\Omega_e}

\newcommand{\winf}{w_\infty}
\newcommand{\ginf}{\gamma_\infty}

\newcommand{\lam}{\Lambda} 
\newcommand{\gppn}{\gamma_{\rm PPN}}

\def\fun#1#2{\lower3.6pt\vbox{\baselineskip0pt\lineskip.9pt
  \ialign{$\mathsurround=0pt#1\hfil##\hfil$\crcr#2\crcr\sim\crcr}}}

\begin{document} 

\title{Parameterized Beyond-Einstein Growth} 
\author{Eric V.\ Linder and Robert N.\ Cahn} 
\affiliation{Lawrence Berkeley National Laboratory, University of 
California, Berkeley, CA 94720} 

\date{\today} 

\begin{abstract} 
A single parameter, the gravitational growth index $\gamma$, succeeds in 
characterizing the growth of density perturbations in the linear 
regime separately from the effects of the cosmic expansion.  The parameter
is restricted to a very narrow range for models of dark energy obeying 
the laws of general relativity but can take on distinctly different
values in models of beyond-Einstein gravity.  
Motivated by the parameterized post-Newtonian 
(PPN) formalism for testing gravity, we analytically derive and extend 
the gravitational growth index, or Minimal Modified Gravity, approach 
to parameterizing beyond-Einstein cosmology. 
The analytic formalism demonstrates how to apply the growth index 
parameter to 
early dark energy, time-varying gravity, DGP braneworld gravity, and
some scalar-tensor gravity.

\end{abstract} 

\maketitle

\section{Introduction} \label{sec:intro} 

The acceleration of the cosmic expansion points to new physics beyond 
the standard models of particle physics or gravitation.  But the nature 
of this physics 
is not clear.  While current data constraints are consistent with 
Einstein's cosmological constant $\lam$, the uncertainties are still 
substantial.  Even the simplest deviation from this picture, an 
assumed constant effective equation of state $w$, is determined 
to no better than 20\% (at 95\% confidence level) when combining the 
full set of current supernova distance, cosmic microwave background, 
and baryon acoustic oscillation measurements, {\it with\/} systematic 
uncertainties included \cite{bighubble}.  Given that virtually no 
physical explanation beyond $\lam$ predicts a simple, purely constant 
equation of state, it is clear that we cannot claim to have zeroed in 
on Einsteinian physics as the solution. 

As we can look beyond Einstein's cosmological constant, so too we can 
look beyond the framework of Einstein gravity.  To address the 
question of the nature of the new physics requires measuring both the 
expansion history of the universe and the history of structure growth.  
The growth history can provide independent information on the accelerating 
physics in the case when gravity deviates from general relativity 
(e.g.\ \cite{luess,knox,spergel,lingrav}).  As we describe in detail below, 
the structure growth history is conveniently described by a 
parameter $\gamma$, first introduced by \cite{pjep80} and expanded 
upon by \cite{wangs}, within Einstein gravity, to supplement the conventional 
cosmological parameters of the matter density and the equation of state 
of the dark energy.  

The goal of this article is to motivate and explore further the idea 
of a model independent parametrization of the deviations of growth from 
Einstein gravity.  Just as the model independent parametrization of the 
expansion history equation of state $w(z)$ proves valuable even without 
a particular Lagrangian in mind, we use $\gamma$ without the intent 
to adopt a specific gravitational action, though we do show results 
for some specific (toy) models. 
We find that $\gamma$ is confined to a narrow range 
of values near $6/11$, independent of the nature of the dark energy if 
general relativity holds.  In contrast, beyond-Einstein models of gravity 
can lead to rather different values for $\gamma$.  This difference then 
establishes a standard for measurements of the growth of structure 
required to exclude alternate models of gravity. 

Linder \cite{groexp} proposed that this independent 
information, apart from the expansion physics, be parameterized by a 
single gravitational growth index $\gamma$, defined through 
\beq
g(a)=e^{\int_0^a (da'/a')\,[\om(a')^\gamma-1]}, \label{eq:gapx}
\eeq
where $\om(a)$ is the matter density fraction of the total energy 
density as a function of scale factor, $g=\delta/a$ and 
$\delta=\delta\rho_m/\rho_m$ is the linear 
density perturbation in the matter.  (See 
\cite{wangs} for an early approach to the growth index within general 
relativity and tied to the expansion history.) \ In \cite{hl06}, 
this was combined with the expansion history into a full cosmological 
fitting framework called Minimal Modified Gravity (MMG), 
using that the expansion history of the universe, $a(t)$, can be 
phrased in terms of an effective equation of state ratio 
$w(a)=w_0+w_a(1-a)$ shown 
to be extremely successful for describing a wide variety of physics, 
including gravitational modifications \cite{linprl,paths}.  
Comparison of the parameters $\{w_0,w_a,\gamma\}$ fitted to data 
(see examples in \cite{hl06}) could establish statistically significant 
evidence for beyond-Einstein physics and provide information on its origin. 

In \S\ref{sec:growth} we reexamine and derive analytically certain 
important properties of the parameter $\gamma$, which were originally
obtained numerically.  The physical motivation for the growth index
approximation to the growth history is strengthened in \S\ref{sec:form}, and 
evaluated for time varying and early dark energy.  We derive a relationship 
between the growth index and the PPN formalism for gravity theories in 
\S\ref{sec:ppn}, and show the success of the growth index in 
characterizing DGP braneworld gravity and some scalar-tensor 
gravity theories. 
The analytic derivations serve an important role as a foundation for 
understanding both extensions and breakdowns of the growth index.  Of 
course exact calculations of growth within specific gravitational models, 
not treated here, play an important role as well.

\section{Parameterizing Growth} \label{sec:growth} 

Considering linear density perturbations in the matter, $\delta= 
\delta\rho_m/\rho_m$, the equation for their growth within general 
relativity is given by 
\beq
\frac{d^2\delta}{dt^2}+2H(a)\frac{d\delta}{dt}-4\pi \rho_m 
\delta=0, \label{eq:maint}
\eeq
where $H=\dot a/a$ is the Hubble parameter.  This can be rewritten as
\beqa 
\frac{dG}{d\ln a}&+&\left(4+\frac{1}{2}\frac{d\ln H^2}{d\ln a}\right)\,G 
+G^2 \nonumber \\ 
&+&3+\frac{1}{2}\frac{d\ln H^2}{d\ln a}-\frac{3}{2}\om(a)=0, \label{eq:main}
\eeqa 
where $G=d\ln(\delta/a)/d\ln a$.  Note that by 
normalizing $\delta$ by $a$ we remove the pure matter universe growth 
behavior, which would give $G=0$.  The dimensionless matter density is
$\om(a)=\om a^{-3}/[H(a)/H_0]^2$, where $\om$ is its present value. 

The growth history can be solved formally by quadrature, giving 
the solution 
\beqa 
G(a)=-1&+&[a^4H(a)]^{-1}\int_0^a \frac{da'}{a'}\,a'^4H(a') \nonumber \\ 
&\times&\left[1+\frac{3}{2}\om(a')-G^2(a')\right]. \label{eq:gaho}
\eeqa 
This can also be written in terms of the dark energy density 
$\Omega_w(a)=1-\om(a)$, for a flat universe, as 
\beqa 
G(a)=-1&+&[a^4H(a)]^{-1}\int_0^a \frac{da'}{a'}\,a'^4H(a') \nonumber \\ 
&\times&\left[\frac{5}{2}-\frac{3}{2}\ow(a')-G^2(a')\right]. \label{eq:gahw}
\eeqa 

For growth during the matter-dominated era, $G$ will be small, 
and this holds reasonably well even as dark energy comes to dominate, since 
$\om$ is still non-negligible today:  for the concordance cosmology, 
 today $G$ is of order $-1/2$.  So let us neglect the $G^2$ term in 
Eqs.~(\ref{eq:gaho})-(\ref{eq:gahw}) -- while this is motivated by the 
asymptotic growth behavior during matter domination, it should be a 
reasonable approximation throughout the growth history.  In practice 
we find that $g(a)=\delta/a$ agrees with the exact solution to $\sim0.2\%$.  
This now gives an explicit solution for the growth. 

We can use 
\beq 
H^2/H_0^2=\om a^{-3}[1+\ow(a)/\om(a)]
\eeq 
to write at the same level of approximation (that the universe is not too far 
from matter-dominated),
\beq 
G(a)=-\frac{1}{2}\ow(a)-\frac{1}{4}a^{-5/2}\int_0^a \frac{da'}{a'}\,a'^{5/2} 
\ow(a'). \label{eq:gow} 
\eeq 
For any particular model of $H(a)$, or $\om(a)$ or $\ow(a)$, we can 
then evaluate the growth history.

Connecting this new analytic approach to the growth index defined in 
Eq.~(\ref{eq:gapx}), we find immediately
\beq 
G(a)=\om(a)^\gamma-1.
\eeq 
In the same approximation used to derive the quadratures for $G$, 
we can write 
\beq 
\gamma\approx-G/\ow(a), \label{eq:gamgow}
\eeq 
and then using Eq.~(\ref{eq:gow}) we have 
\beqa 
\gamma&\approx&\frac{1}{2}+\frac{1}{4}a^{-5/2}\ow(a)^{-1} 
\int_0^a \frac{da'}{a'}\,a'^{5/2}\ow(a') \nonumber \\ 
&=&\frac{1}{2}+\frac{1}{4}\int_0^1 \frac{du}{u}\,u^{5/2} 
\ow(au)/\ow(a). \label{eq:gamw}
\eeqa 

To first order in deviations from matter domination, e.g.\ at early 
times, $\ow(a)\sim a^{-3\winf}$, where $\winf=w(a\ll1)$, and we can 
evaluate Eq.~(\ref{eq:gamw}) to obtain 
\beq 
\gamma_\infty=\frac{3(1-\winf)}{5-6\winf}. \label{eq:ginfw}
\eeq 
Equivalently, we can substitute Eq.~(\ref{eq:gamgow}) into 
Eq.~(\ref{eq:main}) and find 
\beq 
\gamma_\infty=\frac{3+\bar G}{5+2\bar G}, 
\eeq 
where $\bar G\equiv d\ln G/d\ln a=-3\winf$ to this level of approximation. 

These asymptotic values accord exceedingly well with the fitting formula 
of \cite{groexp}.  We can write 
\beq 
\ginf=\frac{6-3(1+\winf)}{11-6(1+\winf)}\approx 
\frac{6}{11}+\frac{3}{121}(1+\winf). \label{eq:ginf1w} 
\eeq 
We can compare this to the fitting formula from \cite{groexp} that gives 
\beqa 
\gamma&=&0.55+0.02\,[1+w(z=1)], \qquad w<-1 \label{eq:gamfita} \\ 
&=&0.55+0.05\,[1+w(z=1)], \qquad w>-1  \label{eq:gamfitb} 
\eeqa 
over the whole range $0<a<1$.  
For the cosmological constant case, $w=-1$, the asymptotic value is 
$\ginf=6/11=0.545$ compared to the numerically obtained fit 
$\gamma=0.55$ for the whole growth history.  

The first order correction to the cosmological 
constant case for a different value of constant $w$ is quite small 
for the asymptotic formula (\ref{eq:ginf1w}), 
with a coefficient of 0.025 times $1+w$.  This agrees with the fitting 
correction from Eqs.~(\ref{eq:gamfita})-(\ref{eq:gamfitb}), which have 
a coefficient of 0.02 (0.05) 
times $1+w$ for the case when $1+w<$ ($>$) 0.  The asymptotic value 
is closer to the fit for the $w<-1$ case because then matter domination 
lasts longer.  The asymptotic form is also in agreement with the 
pioneering work of \cite{wangs}, who expanded the growth equation 
about $\om(a)=1$ for constant $w$.  Their asymptotic term agrees with 
Eq.~(\ref{eq:ginfw}) and they find the next order is given by 
$(15/1331)[1-\om(a)]$ for the $w=-1$ case; even when $1-\om(a)$ is 
much larger than the $\sim 10^{-9}$ for $\Lambda$CDM at CMB last 
scattering, this correction is negligible. 

The new elements in Eq.~(\ref{eq:gamw}) include the integral nature of 
the relation, allowing for treatment of the whole growth history to 
some redshift, rather than an instantaneous measure of growth (thus 
allowing a prediction of $G_{\rm today}$ unlike previous work), the 
ability to treat time varying dark energy, and the clear identification 
of the key assumptions such as early matter domination that allow the 
single parameter to succeed. 

We emphasize that the numerical fitting form of \cite{groexp} covers the 
entire growth history, not just the asymptotic high redshift behavior, and 
does not assume $G\ll1$ but rather fits to the exact numerical solution.  
Nevertheless, it is instructive to pursue the analytic arguments further 
to understand the motivation for phrasing the growth history in terms of 
a gravitational growth index, and why a single parameter proves so 
successful, even in the case of time varying equations of state.

\section{Growth and Dark Energy} \label{sec:form} 

One question to ask is why the fitting form (\ref{eq:gapx}) is 
physically appropriate.  In the asymptotic limit, $G$ will be linearly 
proportional to $\ow(a)=1-\om(a)$ since Eq.~(\ref{eq:main}) is linear 
(for small $G$; cf.\ Eq.~\ref{eq:gow}), 
but one can imagine other forms besides $\om(a)^\gamma-1$ that have 
this property.  We consider three possibilities for fitting forms of $G$: 
the standard one of $\om(a)^\gamma-1$, one directly proportional to 
$\ow(a)$, and $\ln\om(a)^\gamma$.  Each of these has the appropriate limit 
that $G$ vanishes linearly in  $1-\om(a)$ as  $a\to0$.  Defining $\gamma$ for the entire 
growth history in terms of these forms gives the options 
\beq 
\gamma \equiv \frac{\ln (1+G)}{\ln\om(a)},\quad -\frac{G}{\ow(a)},\quad 
\frac{G}{\ln\om(a)}. \label{eq:gamopt}
\eeq 

As a guide to defining $\gamma$ as a useful parameter (e.g.\ nearly constant) 
over the entire growth history from $a=0$ to the present, we can 
examine its late time behavior, when $a\gg1$.  For dark energy 
domination in the future, the matter density $\om(a)\to0$ and 
matter density perturbations $\delta$ stop growing, so $G\to -1$, 
in agreement with Eq. (\ref{eq:gaho}).  The 
second definition of $\gamma$ in Eq.~(\ref{eq:gamopt}) leads 
to $\gamma\to 1$ in the future; the third definition has $\gamma\to0$. 
Only the first, original definition of the growth index preserves its 
stability over the entire growth history from asymptotic past to 
asymptotic future.  Indeed, $\gamma$ varies from its present value by 
less than 2\% back to $a\ll1$ and 6\% forward to $a=5$, for the 
cosmological constant case.  Since $\gamma$ enters into an integral relation, 
then even this mild variation is further smoothed over, giving a 
constant $\gamma$ as an excellent approximation.  \cite{groexp} found 
it to reproduce the exact growth for a wide variety of models to 
better than 0.2\%.  
Overall, the definition of a gravitational growth index through 
Eq.~(\ref{eq:gapx}) is therefore physically well motivated. 

We can carry this further, showing that a single growth parameter suffices 
rather than needing a function (of redshift).  Considering dynamical dark 
energy, rather than a constant $w$ model, expand 
\beq 
\ow(a)\sim e^{3\int_a^1 (da'/a')\,w(a')}\sim a^{-3\winf}\,(1+Ba^x), 
\eeq 
in Eq.~(\ref{eq:gamw}), where we approximate the time variation of the 
equation of state at high redshift  as a power law correction to the 
asymptotic $\winf$ value.  The solution for the growth index is 
\beq 
\gamma\approx\frac{3(1-\winf)}{5-6\winf}-\frac{Bx\,a^x}{(5-6\winf)(5-6\winf 
+2x)}. 
\eeq 
For the standard equation of state parameterization 
$w(a)=w_0+w_a(1-a)=\winf-w_a a$, 
we have $B=3w_a$, $x=1$.  The correction to $\gamma$ is exceedingly small, 
at early times proportional to $a$, with furthermore a small coefficient 
$-(3/143)w_a$, generally $\lesssim0.01$.  So even in the dynamical dark 
energy case, $\gamma$ is found to be nearly constant.  (The growth index 
fit was examined numerically in \cite{hl06}.) 

Suppose we consider early dark energy, where its energy density is 
not completely negligible at high redshifts, but possibly contributing 
up to a couple percent of the total energy density at CMB last scattering 
\cite{wett,early,doranrobbers}.  We might expect this to upset the 
gravitational growth index since early dark energy directly affects the 
matter domination used to derive the analytic asymptotic behavior. 
In order for dark energy to contribute non-negligibly to the early 
energy density, its equation of state must approach $w=0$.  
Substituting $\winf=0$ 
in Eq.~(\ref{eq:ginfw}), we see $\ginf=3/5$.  However, we can actually 
obtain this solution without approximating $G\ll1$; the exact solution 
for growth in a universe with two components, each evolving (at early 
times) with density proportional to $a^{-3}$, as ordinary matter, but 
with a fraction $\Omega_e$ not clustering, is \cite{fry,linmpa,early} 
\beq 
g=\delta/a\sim a^{(-5+\sqrt{25-24\ome})/4}\approx a^{-(3/5)\ome}, 
\eeq 
or $\gamma=3/5$.  (To next order, $\gamma\approx 3/5+(3/125)\ome$, in 
agreement with \cite{wangs}.) 

The redshift distortion factor $\beta$ \cite{kaiser} used in galaxy 
redshift surveys can be written as 
$\beta\sim 1+G=\om^\gamma$, and one often sees $\om^{0.6}$ used.  For 
the concordance model we have seen that $\gamma=0.55$ is a much better 
fit.  Using $\om^{0.6}$ introduces a needless 6\% error 
in $\beta$, or a systematic bias of 0.03 (11\%) in the value of $\om$ derived, 
so more accurate results will be achieved if $\om^{0.55}$ is used in 
place of $\om^{0.6}$.

\section{Growth and Gravity} \label{sec:ppn} 

To this point, the growth behavior has been completely specified 
by the expansion history.  That is, in Eq.~(\ref{eq:main}) the 
terms only involve $H(a)$ or $\om(a)$, and so the gravitational 
growth index $\gamma$ can be defined in terms of the effective 
equation of state 
$w(a)$.  Furthermore, we have seen that $\gamma$ has quite a small 
range over a reasonably large variety of dark energy models, i.e.\ 
for $w=-1$ to $-1/3$ (including effective curvature energy), $\gamma$ 
only varies over 0.55-0.57. 

Now we consider what happens when we allow alterations to the gravity 
theory.  There is no unique 
prescription for how modified gravity theories affect the growth equation, 
even in the linear regime, though see \cite{groexp} for an attempt 
to provide a somewhat general, if formal, treatment.  Effects on 
the growth equation include the introduction of scale dependence, 
anisotropic stress, and variation of the gravitational coupling 
(i.e.\ Newton's constant).  We briefly discuss the first two of 
these in \S\S\ref{sec:ppngrow} and \ref{sec:concl}, and concentrate here 
on the last one.  Again we emphasize that our goal is a model independent 
approach, rather than adopting a specific theory. 

\subsection{Varying gravity} \label{sec:varyg} 

The last term of Eq.~(\ref{eq:main}) contains a hidden dependence 
on the gravitational coupling, multiplying the source term $\om(a)$. 
Here we rewrite the growth equation, explicitly showing the effect: 
\beqa 
\frac{dG}{d\ln a}&+&\left(4+\frac{1}{2}\frac{d\ln H^2}{d\ln a}\right)\,G 
+G^2+3+\frac{1}{2}\frac{d\ln H^2}{d\ln a} \nonumber \\ 
&-&\frac{3}{2}[1+(Q(a)-1)]\, \om(a)=0, \label{eq:mainq}
\eeqa 
where $Q-1$ gives the fractional deviation of the coupling from 
the general relativistic case (i.e.\ $Q=1$ means the coupling is given 
by Newton's constant). 

We can now repeat the analysis of \S\ref{sec:growth}, finding 
\beq 
G(a)=-1+(a^4H)^{-1}\int_0^a \frac{da'}{a'}\,a'^4H\,
\left[1+\frac{3}{2}Q(a')\,\om(a')\right]. 
\eeq 
This is turn leads to a revised gravitational growth index 
\beqa 
\gamma\approx\frac{1}{2}&+&\frac{1}{4}
\int_0^1 \frac{du}{u}\,u^{5/2}\ow(au)/\ow(a) \nonumber \\ 
&-&\frac{3}{2}
\int_0^1 \frac{du}{u}\,u^{5/2}\,[Q(au)-1]/\ow(a). \label{eq:gamqw}
\eeqa 

There are three cases to consider for the gravitational deviation. 
If $(Q-1)\sim a^q$ at early times, and $q>-3\winf$, then the modification 
to $\gamma$ from the altered gravitational coupling will be negligible. 
However, if $q<-3\winf$ then there will be strong modification.  In 
fact, in this case the usual matter-dominated growth behavior at high 
redshift is broken ($G=0$, or $\delta\sim a$, is no longer a solution 
at early times), and large scale structure would not accord with 
observations.  This leaves the main, and physically best motivated, case 
of the scaling behavior $q=-3\winf$ where the same physics responsible for 
the variation in the force of gravity 
also affects the expansion rate, giving rise to an effective energy 
density $\ow(a)$.  We discuss this further in the specific DGP braneworld 
and scalar-tensor gravity examples below. 

In the scaling case, we can write $Q-1=A\ow(a)$ and find the asymptotic 
growth index to be 
\beq 
\ginf=\frac{3(1-\winf-A)}{5-6\winf}. \label{eq:ginfq}
\eeq 
We have verified that, as in \S\ref{sec:form}, it is best to define 
$\gamma$ in the usual way through $G=\om(a)^\gamma-1$, not 
$G=[Q\om(a)]^\gamma-1$, even in the 
presence of beyond-Einstein gravity leading to $Q\ne1$.  That is, 
the gravitational deviation $Q-1$ enters strictly through the growth 
index $\gamma$, while the expansion history determines the growth 
history $G$ apart from the value of $\gamma$.  This separation of physical 
effects from the expansion rate and from the gravity theory through distinct 
beyond-Einstein parameterizations $w$ and $\gamma$ is an important point 
that clarifies the nature of the beyond-Einstein physics.

\subsection{Braneworld gravity} \label{sec:dgp} 

To test our new expression for the gravitational growth index we 
consider several examples of gravity beyond general relativity. 
First, we examine a theory altering the Einstein-Hilbert action 
by a term arising from large extra dimensions, the DGP braneworld 
theory \cite{dgp,ddg}.  On scales smaller than the Hubble scale but 
still within the linear density perturbation regime, the effect on 
the growth behavior is that of a variation in gravitational 
coupling, with \cite{luess}
\beq 
(Q-1)_{\rm DGP}=-\frac{1}{3}\left(\frac{1-\om^2(a)}{1+\om^2(a)}\right). 
\eeq 
Note that, as predicted, one has the scaling behavior $(Q-1)\sim a^{-3\winf}$ 
asymptotically and 
\beq 
A\equiv\frac{Q-1}{1-\om(a)}=-\frac{1}{3}\,\frac{1+\om(a)}{1+\om^2(a)}\,\to  
-\frac{1}{3}, \label{eq:adgp} 
\eeq 
is of order unity.  This ensures that the gravitational deviation neither 
violates matter domination nor has negligible effect on growth. 

The effective equation of state from DGP gravity is $w(a)=-1/[1+\om(a)]$ 
\cite{luess}, approaching $\winf=-1/2$ at high redshift.  Substituting 
the DGP values for $A$ and $\winf$ into Eq.~(\ref{eq:ginfq}), we find 
the growth index 
\beq 
\gamma_{\infty,{\rm DGP}}=\frac{11}{16}=0.6875. 
\eeq 
This accords exactly with the asymptotic numerical solution and extremely 
well with the gravitational growth index fit $\gamma=0.68$ over the whole 
history, given by \cite{groexp}. 

If for any point in the growth history we naively substitute into 
Eq.~(\ref{eq:ginfq}) the values of $A$ from Eq.~(\ref{eq:adgp}) and 
$w(a)=-1/[1+\om(a)]$, we obtain 
\beq 
\gamma_{\rm DGP}\approx\frac{7+5\om(a)+7\om^2(a)+3\om^3(a)}{[1+\om^2(a)] 
\,[11+5\om(a)]}, \label{eq:gamdgp}
\eeq 
which agrees well with the exact numerical solution. 
The growth index takes on the value $11/16=0.6875$ in the asymptotic past 
and 0.634 in the asymptotic future (vs.\ $7/11=0.636$ from the formula, 
which was derived under matter domination). 
At the present the numerical value is 0.665 (for $\om=0.28$, vs.\ 0.674 
from the formula), so the single 
parameter $\gamma=0.68$ is an excellent approximation for 
growth at any time through the present, holding constant to 2\%. 
In addition to the 
remarkable constancy of the growth index, $\gamma$ stays well distinct of 
the pure expansion history prediction within general relativity of 
$\gamma=0.55-0.56$ as the DGP equation of state evolves from $w=-1/2$ in 
the past to $-1$ in the future.  

\begin{figure}[!htb]
\begin{center} 
\psfig{file=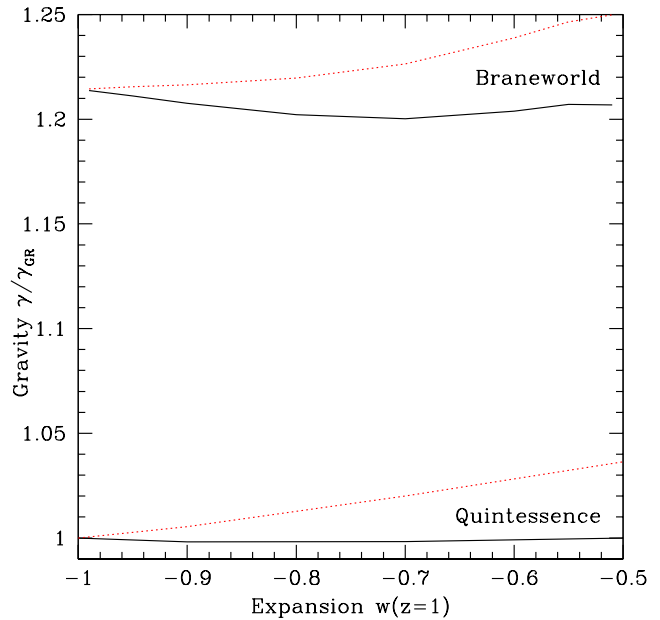,width=3.4in} 
\caption{In the parameterized beyond-Einstein approach, the growth 
index $\gamma$ tests gravity separately from the expansion and 
clearly distinguishes beyond-Einstein gravity from physical dark energy. 
The curves show the values of $\gamma$ change little as the expansion history, 
here represented by the equation of state $w(z=1)$, varies.  
Solid curves show the ratio of the numerical solution for $\gamma$ in the 
braneworld case to that in general relativity (GR), and in the 
quintessence case the ratio of 
the fitting form $\gamma=0.55+0.04\, [1+w(z=1)]$ to the numerical GR 
solution.  (Dotted curves assume $\gamma=0.55$, the GR value in a 
cosmological constant model.)  
}
\label{fig:gammagr} 
\end{center} 
\end{figure}

We illustrate this important property of separation of growth history 
parametrization from expansion history in Figure~\ref{fig:gammagr}.  Here 
the growth index is shown as a ratio to the (exact) general relativity (GR) 
value for the cases of quintessence and braneworld models.  
For the braneworld case, being a single parameter model, $w(z=1)$ determines 
$\om$ (note $\om$ will be far from 0.28 when $w(z=1)$ is far from $-0.62$). 
We see that the curve of $\gamma$ as we vary the expansion history parameter 
$w(z=1)$ is quite flat, showing success in obtaining a growth parameter 
distinct from expansion effects. 

For the quintessence case (taking $\om=0.28$, $w_0=-1$, and then $w(z=1)$ 
serves as a proxy for $w_a$) the curve representing the 
fitting form Eq.~(\ref{eq:gamfitb}) (with a coefficient 0.04 rather than 
0.05 as a better fit over the restricted range $w(z=1)\in[-1,-0.5]$) is 
within 0.2\% of unity, showing the success of this fitting form. 

The growth index formalism thus possesses these important properties: 1) the 
constancy of $\gamma$ in redshift, allowing a single parameter description 
of the gravity deviations beyond-Einstein, 2) the independence of $\gamma$ 
from the expansion history, i.e.\ separating out the expansion 
influence on the growth so as to give a distinct window on the 
gravitational physics, 
and 3) clear signal of the origin of the beyond-Einstein physics, 
achieving in the braneworld case over 20\% deviation from the general 
relativity prediction -- while the ``noise'' from the expansion 
history $w$ within general relativity affects $\gamma$ at the 0.2\% 
level, a factor of 100 less.  
Thus, the gravitational growth index provides a clear and effective 
parameterization of beyond-Einstein gravity. 
(While the theory signal-to-noise is high, achieving observational 
constraints is more challenging, with \cite{hl06} estimating that 
next generation experiments will determine $\gamma$ to within 8\%.)

\subsection{Scalar-tensor gravity} \label{sec:stgrav} 

Next we consider scalar-tensor theories of gravity, involving 
coupling of a scalar field $\phi$ to the Ricci curvature $R$ of 
the form of $F(\phi)\,R$ in the action.  Because of 
the physical coupling between the modification of the expansion and the 
modification of the growth equation, we again might expect the scaling 
relation to hold, where the gravitational deviation $Q-1$ is of 
order the effective energy density $\ow(a)$ in the expansion, 
at least for theories consistent with observations of large scale structure. 
This would give an appreciable, but not pathological, influence on growth. 
Scalar-tensor theories where the scalar field is responsible for the current 
acceleration are often called extended quintessence theories \cite{eq}. 

In extended quintessence (EQ), one can show that the main effects on 
expansion history come from the scalar potential (self-interaction) 
and the change in the gravitational coupling $(8\pi G_N)^{-1}\to F$, 
where $G_N$ is Newton's constant. 
For the growth equation, deviations arise primarily from the variation 
of the coupling, with additional terms being 
suppressed by factors of $(k/H)^{-2}$, where $k$ is the wavemode of 
the density perturbation.  For calculations of $w(a)$ and the 
modified Poisson equation, see \cite{bmpl}. 

The effective modification to the growth history can be treated through 
a deviation in the source term, with $Q-1=(8\pi G_N F)^{-1}-1$.  Given 
a form for $F$, this can be used in Eq.~(\ref{eq:ginfq}) to obtain the 
gravitational growth index for the scalar-tensor theory.  In EQ, the 
$R$-boost mechanism \cite{bmp} operating during matter domination drives 
the theory toward general relativity, so we have a consistent picture 
of the usual matter-dominated growth around the time of CMB last scattering. 
As the scalar field comes to dominate in the late universe, the 
theory may diverge from Einstein gravity.  (Because we are exploring 
the growth index formalism, we do not here worry about constraints from 
solar system tests of gravity or about specific scalar-tensor theories.)  

To give some flavor of calculating $\gamma$, we adopt a toy model with 
coupling 
\beq 
F=\frac{1}{8\pi G_N}\left[1+\frac{B}{1+(a/a_*)^{-q}}\right], 
\eeq 
where $B$ is the amplitude of the coupling variation and $a_*$ is a 
transition scale factor.  At early times, $(Q-1)\sim(a/a_*)^q$. 
The dark energy density (the quintessence part 
of EQ) varies as $\ow(a)\sim a^{-3\winf}$.  As before, 
if $q>-3\winf$, the growth deviation arising from $(Q-1)/\ow(a)$ in 
Eq.~(\ref{eq:ginfq}) will be negligible; if $q<-3\winf$ then the 
growth source term will be drastically affected at early times. 
As mentioned before, the coupling between the scalar field evolution, 
and hence gravity deviation, and the expansion provides a motivation 
for the scaling behavior $q=-3\winf$.  In this regime, 
\beq 
A\equiv\frac{Q-1}{\ow(a)}=-B\,\frac{\om}{1-\om}\,a_*^{3\winf}. 
\eeq 
Substituting this into Eq.~(\ref{eq:ginfq}) predicts $\gamma=0.571$ 
for $a_*=0.5$, $B=0.03$, and $\winf=-1$.  Numerical solution of 
the growth equation 
gives $\gamma=0.571$ at high redshift, $\gamma=0.564$ today.  
This is close to the Einstein range since $B$ is small, but the 
main point is that $\gamma$ is quite constant over the growth history.  
However, for parameter values such that the gravity deviation causes $A$ 
to approach unity, 
$\gamma$ does start to vary (but of course there would be severe 
departures from general relativity today).  
As mentioned, this was a toy model giving but a brief taste of the 
rich phenomenology of scalar-tensor gravity and a full analysis should take 
into account all the observational constraints.  (Also see \cite{bean} 
for analysis of linear perturbations within $f(R)$ gravity.)

\subsection{Relation to PPN} \label{sec:ppngrow} 

As we survey a larger range of gravitational theories, the situation 
becomes more complex. 
The matter source term in the growth equation does not arise purely 
from the Newton-Poisson equation $\nabla^2\Phi_N=4\pi\delta\rho$, 
or $\Phi_N=-4\pi k^{-2}\rho a^3(\delta/a)$ in Fourier space, where 
$\Phi_N$ is the Newtonian potential.  The equations of motion 
actually depend on two potentials, $\Phi$ and $\Psi$, appearing 
in the metric as 
\beq 
ds^2=-(1+2\Psi)dt^2+a^2(t)\,(1+2\Phi)d\vec x^2, 
\eeq 
(written for a flat universe and in longitudinal gauge for simplicity).  

The second order differential equation for the density contrast, {\`a} 
la Eq.~(\ref{eq:maint}), comes from both the density and velocity 
perturbation evolution, and involves both $\Psi$ and $\Phi$.  The 
combination 
\beq 
\Psi+\Phi=-k^{-2}\Pi,
\eeq 
is not necessarily zero, as in general relativity, in the presence of 
anisotropic stress $\Pi$.  
While saying anything general about growth is difficult, let us 
motivate one approach. 

The fractional correction to the source term was defined as $Q-1$ in
\S\ref{sec:varyg}.  
For wavemodes on sub-Hubble scales, the 
dominant correction to the growth equation takes the form (following 
the derivations of \cite{bardeen,kodamasasaki}) 
\beq 
-k^2\Psi \to -k^2\Psi + (2/3)\,k^4(\Psi+\Phi)/\rho_m. 
\eeq 
We note that we expect the scaling behavior 
to hold because the anisotropic stress is generated by the dark 
energy fluid.  The deviation $Q-1$, and the factor $A$ in 
Eq.~(\ref{eq:ginfq}), is proportional to $1+\Phi/\Psi$. 
We can now consider a direct connection to the parameterized post-Newtonian 
(PPN) formalism \cite{will}.  The first PPN parameter is just the ratio of the 
first order corrections of the $g_{ii}$ and $g_{00}$ parts of the metric, 
hence equivalent to $-\Phi/\Psi$.  This PPN parameter is unfortunately 
also given the symbol $\gamma$, so we will denote it as $\gamma_{\rm PPN}$. 
Following \S\ref{sec:varyg}, we see a correction to the growth index, 
\beq 
\gamma=\frac{3(1-\winf+(2/9)(k/H)^2[1-\gppn])}{5-6\winf}. 
\eeq 
Note that in this case $\gamma$ has a scale dependence.  (For large 
wavemodes $k$, or small wavelengths, the $k^2$ behavior is cut off 
by other terms we neglected, such as the usual Jeans term of the 
matter pressure.)   As the beyond-Einstein deviation $1-\gppn$ vanishes, 
the growth index approaches the general relativity expression. 
This illustrates but the barest outline of the complexity of 
beyond-Einstein gravity.

\section{Conclusion} \label{sec:concl} 

Observations, while broadly consistent with the concordance cosmology 
of Einstein's cosmological constant within Einstein's general relativity, 
still offer significant leeway for beyond-Einstein physics.  Whether there 
is such a deviation and how to distinguish its physical origin -- from 
a new field or an extended theory of gravity -- are major questions to 
address with the next generation of experiments.  However on the theory 
side, there is no universally accepted model beyond-Einstein.  This 
drives us to develop a usable, model independent formalism, along the 
lines of the parameterized post-Newtonian approach, in which to evaluate 
future data and guide us to a theory. 

Here we have demonstrated that the Minimal Modified Gravity (MMG) 
approach of combining a gravitational growth index $\gamma$ to 
describe the mass perturbation growth history with equation of 
state $w(a)$ parameterization to describe the expansion history is 
in many cases robust, accurate, and broadly applicable.  The growth 
index $\gamma$ 
clarifies the nature of beyond-Einstein physics by separating the 
expansion effects on the growth from the gravitational theory effects. 

Rather than needing a full function, the single growth index parameter 
$\gamma$ ({\`a} la PPN) 
is extraordinarily successful:  1) it reproduces the exact growth history 
in a highly accurate manner ($<0.2\%$ deviation), 2) stays constant 
to high accuracy, 3) describes both physical dark energy (\S\ref{sec:form}) 
and beyond-Einstein gravity models consistent with observations 
(\S\ref{sec:ppn}), and 4) exhibits clear distinction between these 
different physical origins by deviating as the gravity theory does 
(up to $\sim20\%$ from general relativity) even when the 
expansion histories are identical. 

We emphasize, however, that while MMG is surprisingly broad in its 
physical realism and parameterized beyond-Einstein growth is an intriguing 
formalism for exploring new physics, there is considerable work still 
to do.  This article has dealt only with the linear density perturbation 
regime of growth for modes below the Hubble scale; other cases, in 
particular the nonlinear regime, may not be susceptible to a 
model independent approach.  However, \cite{hl06} suggests first steps 
for addressing the translinear regime, $\delta\sim1$, within MMG; this 
region will be important for weak gravitational lensing probes of 
cosmology.  Modes approaching the Hubble scale can be probed to some 
extent by the integrated Sachs-Wolfe effect, and these observations may 
offer some clue to beyond-Einstein physics (see, e.g., \cite{isw}).  
The Hubble scale, and other features like a Vainshtein or Yukawa scale, 
can introduce scale dependence in the growth as well \cite{sawicki,jain}.  
Microphysics such as a 
dark energy sound speed or anisotropic stress (or coupling 
matter to dark energy, or non-minimal coupling to gravity), can 
confuse the interpretation of the growth history, as outlined by 
\cite{lin0611}.  A specific formal model for this was proposed by 
\cite{kunz}, though the dark energy perturbations become large, possibly 
giving observational difficulties. 

Our understanding of the full range of models beyond-Einstein has a 
long way to go, but Minimal Modified Gravity may provide 
a simple, robust, and broadly valid benchmark for testing future 
observations against new physics (cf.\ the role of minimal supergravity 
for dark 
matter physics).  The beyond-Einstein parameterization $\{w_0,w_a,\gamma\}$ 
provides an accessible and reasonable model independent approach to 
studying the physics of the accelerating universe.

\section*{Acknowledgments} 

This work has been supported in part by the Director, Office of Science,
Department of Energy under grant DE-AC02-05CH11231.


\begin{thebibliography}{99}

\bibitem{bighubble}
  M. Kowalski et al., in preparation 

\bibitem{luess}
  A.\ Lue, R.\ Scoccimarro, G.\ Starkman, Phys.\ Rev.\ D 69, 
  124015 (2004) [astro-ph/0401515] 

\bibitem{lingrav}
  E.V. Linder, Phys. Rev. D 70, 023511 (2004) [astro-ph/0402503]

\bibitem{knox}
  L.\ Knox, Y.-S.\ Song, J.A.\ Tyson, Phys.\ Rev.\ D 74, 023512 (2006) 
  [astro-ph/0503644] 

\bibitem{spergel}
  M.\ Ishak, A.\ Upadhye, D.\ Spergel, Phys. Rev. D 74, 043513 (2006) 
[astro-ph/0507184]

\bibitem{pjep80}
  P.J.E.\ Peebles, {\it Large-Scale Structure of the Universe\/}, 
  (Princeton U.\ Press: 1980) 

\bibitem{wangs}
  L.\ Wang \& P.J.\ Steinhardt, Ap.\ J.\ 508, 483 (1998) [astro-ph/9804015]

\bibitem{groexp}
  E.V.\ Linder, Phys.\ Rev.\ D  72,  043529 (2005) [astro-ph/0507263]

\bibitem{hl06}
  D. Huterer \& E.V. Linder, Phys. Rev. D 75, 023519 (2007) [astro-ph/0608681]

\bibitem{linprl}
  E.V.\ Linder, Phys.\ Rev.\ Lett.\ 90, 091301 (2003) [astro-ph/0208512]

\bibitem{paths}
  E.V.\ Linder, Phys.\ Rev.\ D 73, 063010 (2006) [astro-ph/0601052]

\bibitem{wett}
  C.\ Wetterich, Phys.\ Lett.\ B 594, 17 (2004) [astro-ph/0403289] 

\bibitem{doranrobbers} 
  M.\ Doran \& G.\ Robbers, JCAP 0606, 026 (2006) [astro-ph/0601544]

\bibitem{early}
  E.V. Linder, Astropart. Phys. 26, 102 (2006) [astro-ph/0604280]

\bibitem{fry} 
  J.N.\ Fry, Phys.\ Lett.\ B 158, 211 (1985)

\bibitem{linmpa}
  E.V.\ Linder, {\it Notes on Growth of Structure in a Friedmann Universe\/}, 
  (MPA Internal Research Note: 1988) ; 
  E.V.\ Linder, {\it First Principles of Cosmology\/}, (Addison-Wesley: 1997)

\bibitem{kaiser}
  N. Kaiser, MNRAS 227, 1 (1987) 

\bibitem{dgp}
  G. Dvali, G.\ Gabadadze, M.\ Porrati, Phys.\ Lett.\ B 485, 208 (2000) 
  [hep-th/0005016]

\bibitem{ddg}
  C.\ Deffayet, G.\ Dvali, G.\ Gabadadze, Phys.\ Rev.\ D 65, 044023 (2002) 
  [astro-ph/0105068]

\bibitem{eq}
  F. Perrotta, C. Baccigalupi, S. Matarrese, Phys. Rev. D 61, 023507 (2000) 
  [astro-ph/9906066]

\bibitem{bmpl}
  C.\ Baccigalupi et al., in draft 

\bibitem{bmp}
  C. Baccigalupi, S. Matarrese, F. Perrotta, Phys. Rev. D 62, 123510 (2000) 
  [astro-ph/0005543] 

\bibitem{bean} 
  R. Bean et al., Phys. Rev. D 75, 064020 (2007) [astro-ph/0611321] 

\bibitem{bardeen}
  J.M. Bardeen, Phys. Rev. D 22, 1882 (1980) 

\bibitem{kodamasasaki}
  H. Kodama \& M. Sasaki, Prog. Th. Phys. Suppl. 78, 1 (1984) 

\bibitem{will}
  C.M. Will, Living Rev. Relativity 9, 3 (2006)

\bibitem{isw}
  L.\ Pogosian et al., Phys. Rev. D 72, 103519 (2005) [astro-ph/0506396]

\bibitem{sawicki}
  I.\ Sawicki, Y.-S.\ Song and W.\ Hu, Phys. Rev. D 75, 064002 (2007) 
[astro-ph/0606285] 

\bibitem{jain}
  H.F.\ Stabenau and B.\ Jain, Phys. Rev. D 74, 084007 (2006) 
[astro-ph/0604038] 

\bibitem{lin0611} 
E.V. Linder, Journal of Physics A 40, 6697 (2007) [astro-ph/0610173] 

\bibitem{kunz}
  M. Kunz \& D. Sapone, Phys. Rev. Lett. 98, 121301 (2007) [astro-ph/0612452]


\end{thebibliography}
\end{document}